\documentclass[french]{pfia}
\usepackage{cite}
\usepackage{amsmath,amssymb,amsfonts}
\newtheorem{definition}{Definition}
\usepackage{algorithmic}
\usepackage{graphicx}
\usepackage{textcomp}
\usepackage{xcolor}
\usepackage{seqsplit}
\usepackage{tikz}
\usepackage{tabularx}
\usepackage{array}
\usepackage{listings}
\usepackage{adjustbox}
\usepackage{multirow}
\usepackage{color}
\usepackage{seqsplit}
\newcolumntype{P}[1]{>{\centering\arraybackslash}p{#1}}
\newcolumntype{M}[1]{>{\centering\arraybackslash}m{#1}}

\title{\textbf{Système de recommandations basé sur les contraintes pour les simulations de gestion de crise}}

\author{Ngoc Luyen Le\fup{1,2}, Jinfeng Zhong\fup{3}, Elsa Negre\fup{3}, Marie-Hélène Abel\fup{1}\\[6pt]
	\fup{1} Université de technologie de Compiègne, CNRS, Heudiasyc\\(Heuristics and Diagnosis of Complex Systems), CS 60319 - 60203 Compiègne Cedex, France.\\
	\fup{2} Vivocaz, 8 B Rue de la Gare, 02200, Mercin-et-Vaux, France.\\
	\fup{3} Université Paris Dauphine-PSL, Université Paris Sciences et Lettres,\\ CNRS UMR 7243, LAMSADE, Paris, France.}

\date{}

\begin{document}
	
	\maketitle
	
	
	\begin{resume}
		Dans le cadre de l'évacuation des populations, certains citoyens/bénévoles peuvent et souhaitent participer à l'évacuation des populations en difficulté en venant prêter main-forte aux véhicules d'urgence/évacuation avec leurs propres véhicules. Une manière de cadrer ces élans de solidarité serait de pouvoir répertorier en temps réel les citoyens/bénévoles disponibles avec leurs véhicules (terrestres, maritimes, aériennes, etc.), de pouvoir les géolocaliser en fonction des zones à risques à évacuer et de les ajouter aux véhicules d'évacuation. Parce qu'il est difficile de proposer un système opérationnel temps réel efficace sur le terrain en situation de crise réelle, nous proposons dans ce travail d'ajouter un module de recommandation de couples conducteur/véhicule (avec leurs spécificités) à un système de simulation de gestion de crise. Pour ce faire, nous avons choisi de modéliser et de développer un système de recommandations basé sur des contraintes s'appuyant sur une ontologie pour les simulations de gestion de crise.
	\end{resume}
	
	\begin{motscles}
		Graphe de connaissances, Système de recommandations, Ontologie, Simulations, Gestion de crise
	\end{motscles}
	
	\begin{abstract}
		In the context of the evacuation of populations, some citizens/volunteers may want and be able to participate in the evacuation of populations in difficulty by coming to lend a hand to emergency/evacuation vehicles with their own vehicles. One way of framing these impulses of solidarity would be to be able to list in real-time the citizens/volunteers available with their vehicles (land, sea, air, etc.), to be able to geolocate them according to the risk areas to be evacuated, and adding them to the evacuation/rescue vehicles. Because it is difficult to propose an effective real-time operational system on the field in a real crisis situation, in this work, we propose to add a module for recommending driver/vehicle pairs (with their specificities) to a system of crisis management simulation. To do that, we chose to model and develop an ontology-supported constraint-based recommender system for crisis management simulations.
	\end{abstract}
	
	
	
	\section{Introduction}
	
	Dans le contexte de l'évacuation des populations, les ressources publiques traditionnelles telles que les ambulances et les hélicoptères de gendarmerie (avec des conducteurs professionnels) peuvent être limitées et mal positionnées pour atteindre toutes les personnes dans le besoin. Dans de telles situations, il est nécessaire d'explorer des ressources d'évacuation alternatives. Les ressources citoyennes, en revanche, sont généralement plus dispersées et donc plus accessibles. De plus, de nombreux citoyens/bénévoles\footnote{Dans cet article, nous nous concentrons sur les conducteurs citoyens/bénévoles (parmi les citoyens qui ne sont pas impliqués dans la gestion de crise), ainsi que sur leurs propres véhicules, que nous résumerons par le terme de ``conducteurs/véhicules citoyens/bénévoles''.}  peuvent être disposés à aider à l'évacuation en utilisant leurs propres véhicules. Par exemple, un propriétaire de minibus avec une capacité de 9 passagers pourrait potentiellement évacuer 8 personnes supplémentaires, augmentant considérablement la capacité du processus d'évacuation. De même, un propriétaire de bateau avec une capacité de 6 passagers pourrait aider à évacuer 5 personnes en cas d'inondation. Malheureusement, les recherches existantes sur la gestion de crise (par exemple, la simulation d'évacuation) \cite{laatabi2022coupling, bennett2017heuristic, elsergany2015development,jain2018ontology, mehla2020ontology, zhang2016ontology} se sont principalement concentrées sur l'utilisation de ressources publiques telles que les ambulances et les camions de pompiers. Cependant, dans certains cas, ces ressources peuvent ne pas être disponibles en raison d'une demande élevée ou de la localisation éloignée de la zone touchée. De plus, la localisation des ressources publiques peut également être impactée par une crise, aggravant la pénurie de ressources.
	
	En réalisant ce travail, nous apportons les contributions suivantes : (i) nous étudions le développement d'une ontologie pour aider à organiser des vocabulaires partagés, standardiser les connaissances liées à la gestion de crise et faciliter la mise en œuvre d'un système de recommandations basé sur des contraintes. En utilisant cette ontologie, nous pouvons rationaliser le processus de réutilisation des informations afin d'améliorer l'efficacité du système de recommandation basé sur des contraintes; (ii) nous formulons le problème de distribution de véhicules lors de la mise à l'abri des populations comme un problème de recommandation, ce qui nous permet d'incorporer différentes techniques de recommandation pour allouer efficacement les ressources. Plus précisément, notre système de simulation de crise basé sur une ontologie pour la mise à l'abri des populations vise à consolider les ressources citoyennes pour aider à mettre les populations à l'abri lors d'une crise, en particulier dans les situations où les ressources publiques sont insuffisantes.
	
	Le reste de cet article est organisé comme suit. La section suivante présente la construction de l'ontologie et notre modèle de système de recommandations basé sur des contraintes pour les simulations de gestion de crise. Dans la troisième section, nous présentons notre prototype et son application sur un cas d'utilisation détaillé. Enfin, nous concluons en proposant quelques pistes de travaux futurs.
	
	\section{Notre approche}
	\subsection{Formulation du problème}

 Nos travaux visent à modéliser et proposer un système de gestion des ressources conducteur/véhicule pour l'évacuation des populations touchées par une crise. Deux problèmes clés sont abordés : ($P1$) l'organisation des données et informations relatives aux ressources conducteur/véhicule, et ($P2$) la recommandation de solutions optimales en tenant compte des contraintes de capacité, de temps de réponse et de contexte. Le problème ($P1$) se concentre sur le choix d'un modèle approprié pour organiser les données et l'information dans le contexte de la gestion de crise. La modélisation ontologique est utilisée pour capturer et représenter les concepts et les relations du domaine de la gestion de crise. Le problème ($P2$) concerne la conception et le développement d'un système de recommandations capable de proposer des solutions d'allocation de ressources adaptées à chaque situation. Les technologies de recommandation basées sur les connaissances et l'ontologie permettent de prendre en compte les exigences spécifiques des points de secours\footnote[1]{Un point de secours est un lieu spécifiquement désigné, au sein d'une situation de crise, où les personnes peuvent se rendre pour obtenir de l'aide, des soins médicaux ou être évacuées.} et de calculer des recommandations pertinentes en utilisant les connaissances sur le contexte et les ressources disponibles.
	
	\begin{definition}{\vspace{-0.15cm}Un système de recommandations basé sur des contraintes pour l'allocation des ressources} est défini en utilisant 4 ensembles : l'ensemble des ressources mobiles\footnote[2]{Les ressources mobiles se limitent aux ressources citoyennes dans le cadre de notre travail.} $\mathcal{R}$ {avec leurs caractéristiques/attributs}, l'ensemble des points de secours et leurs besoins $\mathcal{P}$, l'ensemble des abris $\mathcal{S}$ {avec leurs caractéristiques/attributs}, et l'ensemble des contraintes $\mathcal{C}$. Une recommandation de solution pertinente est calculée en fonction de l'élément concret, des ensembles $\mathcal{R}$, $\mathcal{P}$, et $\mathcal{S}$ de sorte que les contraintes spécifiées $\mathcal{C}$ soient satisfaites\end{definition}
	

	\begin{definition}
 \vspace{-0.15cm}	{Une tâche de recommandation pour l'allocation des ressources mobiles}
		est définie comme un problème de satisfaction de contraintes $(\mathcal{R}, \mathcal{P}, \mathcal{S}, \mathcal{C})$ basé sur l'attribution et le calcul du nombre de ressources mobiles dans l'ensemble $\mathcal{R}$ allouées à un point de secours dans $\mathcal{P}$ de sorte qu'il satisfait et ne viole aucune des contraintes dans $\mathcal{C}$.\vspace{-0.15cm}
	\end{definition}
	
	Une recommandation de solution optimale pour l'allocation des ressources mobiles dans $\mathcal{R}$ proposera une liste de ressources mobiles disponibles utilisées pour transférer les évacués des points de secours $\mathcal{P}$ vers les abris $\mathcal{S}$ avec un temps de déplacement minimum. Dans la prochaine section, nous détaillerons le développement d'une ontologie en gestion de crise pour les ressources et les facteurs liés.
	
	\subsection{Construction de l'ontologie}
 Dans notre étude, nous avons utilisé la Méthodologie Agile pour le Développement d'Ontologie (AMOD)\cite{abdelghany2019agile}, qui comprend trois phases distinctes : préliminaire, développement et post-développement, chacune contribuant à la réalisation progressive de l'ontologie. Dans la phase préliminaire, l'objectif principal de la construction de l'ontologie est de fournir un modèle standard avec une terminologie et un vocabulaire pour collecter des informations sur les ressources disponibles (par exemple, les véhicules, les conducteurs) qu'une organisation (par exemple, le Conseil Municipal) mobilisera pour évacuer les personnes affectées lors d'une crise. Les concepts centraux de l'ontologie sont définis sur la base des entités importantes avec leurs caractéristiques dans le contexte de la gestion de crise.

\begin{figure}
	\centering
	\includegraphics[width=0.9\linewidth]{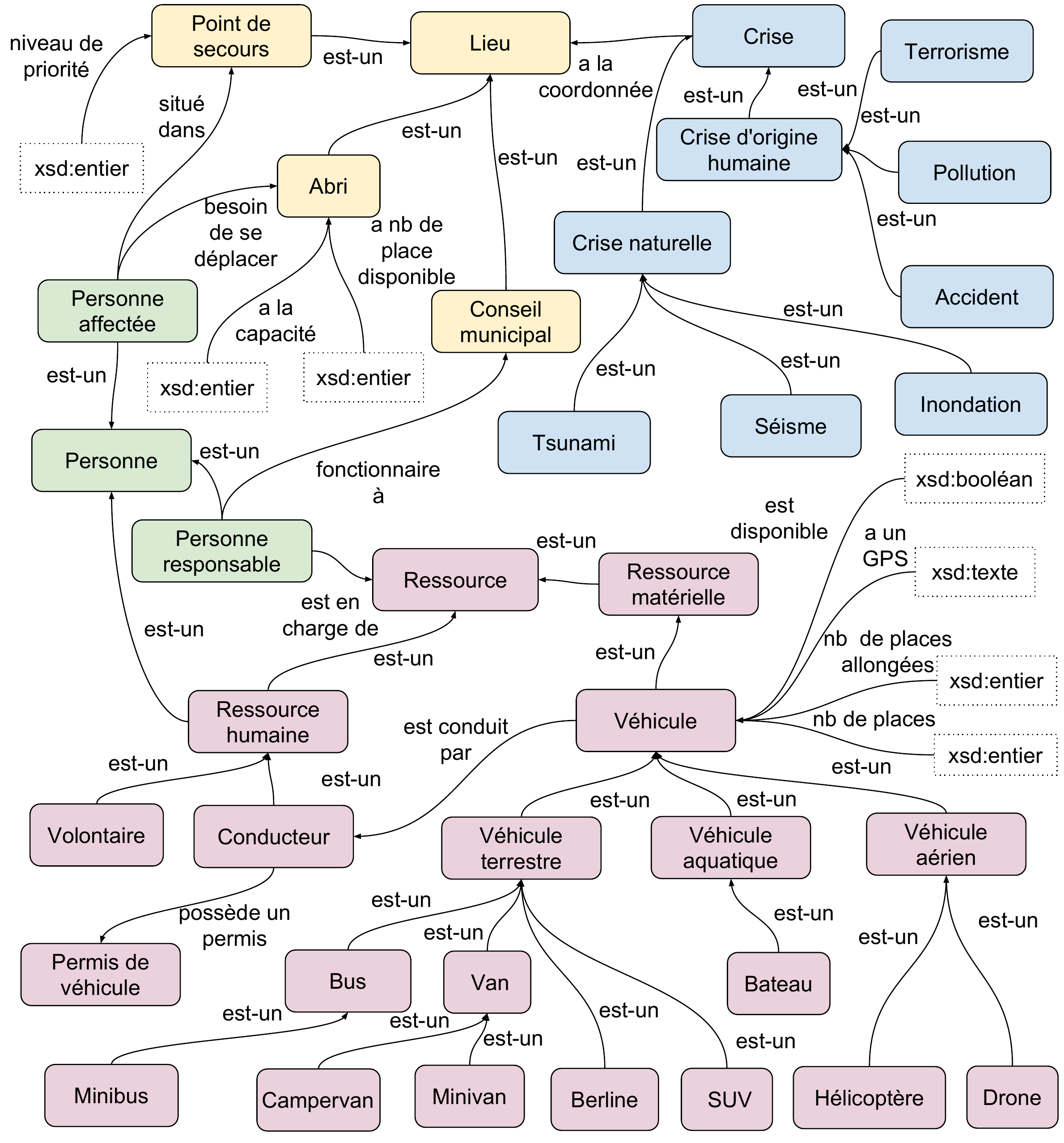}
 \vspace{-0.3cm}
	\caption{Un extrait de l'ontologie concernant les ressources conducteur/véhicule, les lieux et les personnes concernées dans la gestion d'une crise.}
 \vspace{-0.5cm}
	\label{fig_01}
\end{figure}	
	
 Pendant la phase de développement, nous organisons des sprints\footnote[3]{Un sprint est une unité de temps fixe pendant laquelle un ensemble spécifique de tâches doit être accompli.} et choisissons de développer notre ontologie en s'appuyant sur l'ontologie \textit{ISyCri} \cite{benaben2008metamodel} en utilisant des concepts liés à la description de la crise, des personnes touchées et des ressources. 
 Plus précisément, comme illustré par la Figure~\ref{fig_01}, nous adaptons et développons notre modèle ontologique autour de trois principales entités : les ressources, les personnes et les lieux. Premièrement, les ressources sont distinguées selon qu'elles sont humaines, matérielles ou mobiles. Dans notre cas, les ressources humaines sont les citoyens/bénévoles qui participent aux opérations de secours et d'évacuation. Tandis que les ressources matérielles incluent les catégories de véhicules et leurs informations descriptives. Une ressource mobile sera représentée par une association par défaut d'un véhicule et d'un conducteur (c'est-à-dire une paire de ressources conducteur/véhicule). 
	Deuxièmement, l'identification et l'organisation des lieux joue un rôle extrêmement important dans notre cas. Chaque lieu doit être spécifiquement identifié avec des informations de localisation. En général, les lieux sont séparés en points de secours et abris. Les points de secours sont des sites où les personnes affectées se regroupent et sont acheminées vers un abri par une ressource mobile. 
	
	Enfin, les populations peuvent être distinguées entre les populations affectées et les ressources humaines. Les populations affectées sont les populations vulnérables lors de la crise, et elles ont besoin d'être déplacées vers un abri. Tandis que les ressources humaines peuvent être des conducteurs qui utilisent leur véhicule pour participer aux activités d'évacuation. En général, la représentation d'une personne par l'ontologie est utile pour rassembler les ressources humaines et les informations sur les personnes affectées dans les étapes préalables et postérieures à la crise.
		
	\subsection{Système de recommandation}\label{sec:constraint_based}
	Pendant la mise à l'abri des populations, la distribution efficace des conducteurs/véhicules citoyens/bénévoles disponibles vers les zones touchées est un sujet de recherche crucial. Ce problème peut être considéré comme un problème de recommandation, où les conducteurs/véhicules citoyens/bénévoles sont traités comme des éléments à recommander et les points de secours sont traitées comme des utilisateurs auxquels les conducteurs/véhicules citoyens/bénévoles doivent être recommandés. Dans cette section, nous fournissons une description détaillée du système de recommandations basé sur les contraintes pour les simulations de gestion de crise. Les systèmes de recommandations basés sur les contraintes génèrent des recommandations en identifiant les éléments qui répondent à un ensemble de contraintes explicites prédéfinies. Dans notre cas, le système de recommandations vise à générer des paires de conducteurs/véhicules pour chaque point de secours, en veillant à ce que les conducteurs/véhicules affectés à chaque point de secours aient une capacité suffisante pour évacuer la population tout en minimisant le temps nécessaire pour atteindre les points de secours (pour plus de détails \cite{hicss23}).

	
	Lorsque plusieurs solutions sont disponibles, notre algorithme renvoie celle qui utilise moins de véhicules pour réduire le temps total nécessaire et le risque d'embouteillages. Nous calculons $T_{CV-RP}$ avec \emph{OSMNX} \cite{boeing2017osmnx}, un package Python qui permet de télécharger des données géospatiales depuis \emph{OpenStreetMap}: {le système énumère exhaustivement toutes les solutions possibles et sélectionne celle qui utilise le nombre minimal de véhicules pour soulager les embouteillages}. {Pour réduire le temps de calcul requis pour générer la liste de recommandations, nous pré-calculons le temps estimé entre chaque point}. Nous avons essayé \emph{Google Maps API}, et il s'est avéré que son utilisation prenait plus de temps que OpenStreetMap pour calculer le temps estimé entre deux points. Nous avons donc adopté ce dernier. {Plus précisément, nous avons utilisé les outils de recherche opérationnelle (OR-Tools) \cite{kruk2018practical} pour l'optimisation combinatoire, conçus pour trouver la solution optimale à un problème à partir d'un ensemble extrêmement large de solutions possibles. Il convient de noter que les ressources publiques peuvent être considérées comme un cas particulier dans notre contexte. Les véhicules publiques sont généralement garés à des points fixes, tandis que les conducteurs/véhicules sont souvent dispersés. Par conséquent, les véhicules publiques tels que les ambulances peuvent également être intégrés à notre contexte.}

	\section{Prototype et cas d'étude}
 
	Suivant les directives de recherche en science de la conception proposées par \cite{hevner2004design} : La recherche en science de la conception doit produire un artefact viable sous la forme d'un construit, d'un modèle, d'une méthode, ou d'une instanciation. Dans cette section, nous présentons un prototype d'un système qui aide à recommander des ressources conducteur/véhicule pendant une crise. Nous présenterons d'abord l'architecture du système qui met en œuvre le système de recommandations basé sur des contraintes, comment le système fonctionne, puis nous décrivons un scénario pour illustrer l'utilité de notre système.

\begin{figure}[h]
		\centering
		\includegraphics[width=0.8\linewidth]{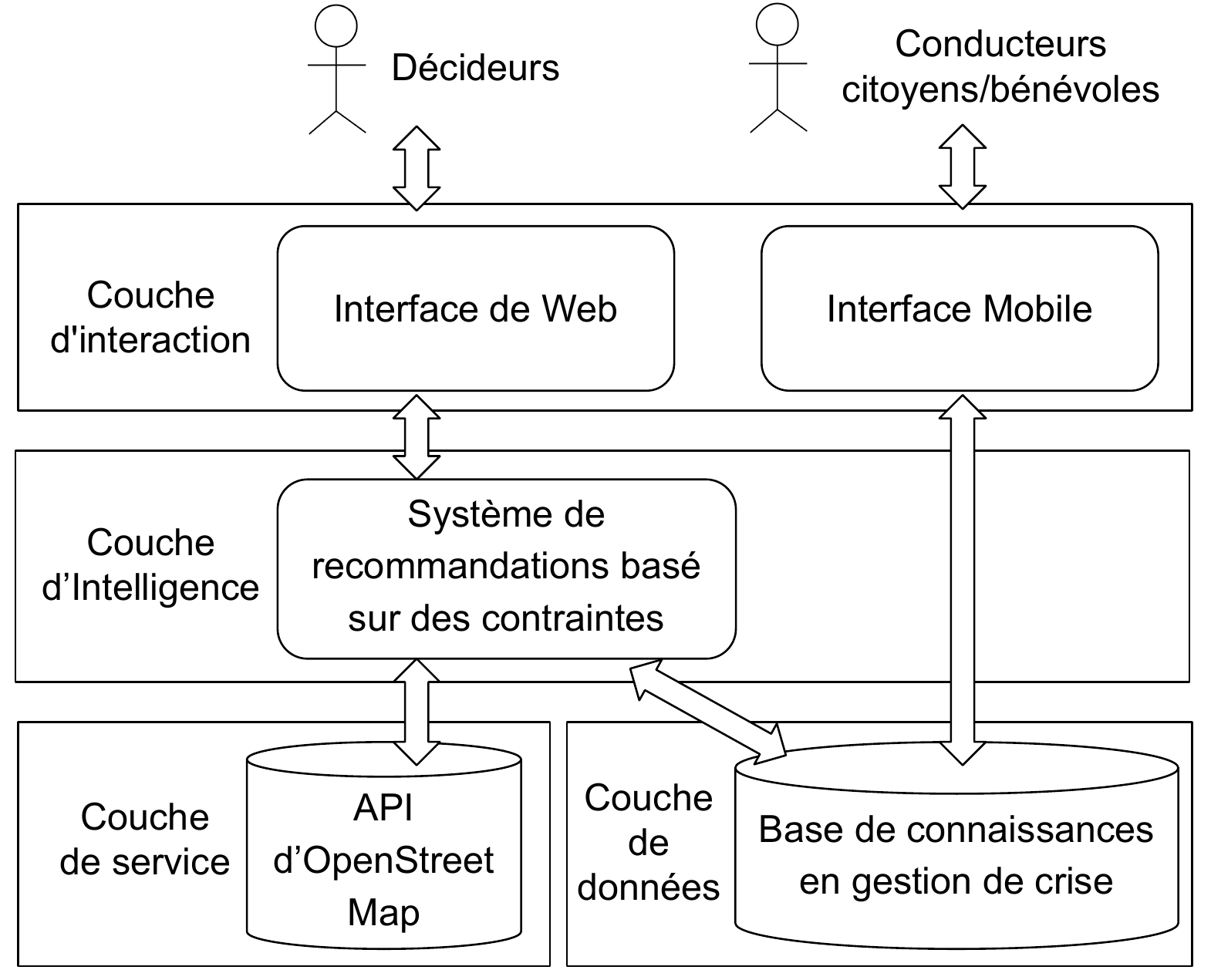}
        \vspace{-0.2cm}
		\caption{L'architecture du système}
		\label{fig:schema}
        \vspace{-0.4cm}
	\end{figure}

	Comme illustré par la Figure~\ref{fig:schema}, 
	notre système est basé sur une architecture à quatre couches. La couche d'interaction comprend une interface mobile qui recueille la disponibilité des conducteurs citoyens/bénévoles pendant une crise, et une interface web permettant aux décideurs d'interagir avec le système et de préciser les informations de chaque point de secours. La couche d'intelligence, le cœur du système, calcule la liste des recommandations satisfaisant certaines contraintes et les affiche aux décideurs. La couche de service calcule les coordonnées de chaque point de secours à partir de sa localisation géographique et estime le temps nécessaire pour que chaque paire conducteur/véhicule atteigne le point de secours. Enfin, la couche de données contient une base de connaissances soutenue par une ontologie pour le domaine de la gestion de crise, modélisant et stockant toutes les informations et données nécessaires.
	A titre d'illustration, prenons une crise d'inondation qui se produit dans la ville de Compiègne et qui nécessite une évacuation rapide des personnes vulnérables vers des abris. Imaginons que le conseil municipal dispose (1) d'une liste de conducteurs citoyens/bénévoles associés et de leur véhicule personnel pouvant être sollicités en situation d'urgence ; (2) d'abris d'accueil d'urgence. A l'aide d'une interface Web, les décideurs devant gérer l'évacuation fournissent au système les informations requises pour chaque point de secours, notamment le nombre de personnes et de personnes handicapées à évacuer, ainsi que leur niveau de priorité. Le système consulte alors sa base de connaissances afin d'identifier les ressources conducteurs/véhicules disponibles. En utilisant les données d'OpenStreetMap, il calcule les temps de déplacement estimés entre les véhicules et les points de secours, en cherchant à les minimiser. Enfin, le système recommande aux  gestionnaires de l'évacuation la liste optimale des paires conducteur/véhicule afin de les aider dans leur prise de décision. 
 Cette approche permet d'optimiser l'allocation des ressources pour évacuer les personnes vulnérables dans les délais les plus courts vers les abris.
	
	\section{Conclusion et perspectives}
	Cet article présente un système de recommandations destiné à aider les décideurs à allouer des paires de conducteur/véhicule citoyens/bénévoles, lorsque les ressources publiques sont insuffisantes. Le système, structuré en quatre couches modulaires interconnectées, utilise une ontologie pour la structuration et le stockage des données, applique OpenStreetMap pour le calcul du temps et de la distance entre deux points géographiques, génère des recommandations pour chaque point de secours et facilite les interactions entre les décideurs et le système de recommandations. Pour l'avenir, nous envisageons d'enrichir l'ontologie pour mieux gérer les crises, d'ajouter davantage de contraintes pour une modélisation plus réaliste de la crise, de construire un système dynamique pour la réutilisation des ressources en temps réel et d'intégrer notre système dans une simulation basée sur des agents pour en évaluer les aspects socio-techniques dans différents scénarios de gestion de crise.
	\section*{Remerciements}
	Cette recherche a été financée par l’Agence National de la Recherche (ANR) et par l’entreprise Vivocaz au titre du projet France Relance – préservation de l’emploi R\&D (ANR-21-PRRD-0072-01).
	
	\bibliographystyle{plain}
	\bibliography{references} 
	
\end{document}